\documentclass{article}
\usepackage[accepted]{mlsys2022}
\usepackage{microtype}
\usepackage{graphicx}
\usepackage{subfigure}
\usepackage{booktabs} 
\usepackage{tabularx}
\usepackage{enumitem}
\usepackage{pifont}
\usepackage{xspace}
\usepackage{color, colortbl}
\usepackage{url}

\usepackage[breaklinks]{hyperref}
\usepackage{breakurl}

\definecolor{LightGray}{gray}{0.99}
\definecolor{LightCyan}{rgb}{0.88,1,1}
\usepackage{tikz}
\usepackage{amsmath}
\usepackage[textsize=tiny,textwidth=0.6in]{todonotes}
\usepackage{setspace}

\def\compactify{\itemsep=2pt \topsep=2pt \partopsep=1pt \parsep=1pt \leftmargin=1.6em}
\let\latexusecounter=\usecounter

\newcommand{\ectwo}{EC2\xspace}
\newcommand{\srift}{Srifty\xspace}

\newcommand{\code}[1]{\texttt{\small{#1}}}
\newcommand{\cmark}{\ding{51}}
\newcommand{\xmark}{\ding{55}}

\mlsystitlerunning{\srift{}: Swift and Thrifty Distributed Training on the Cloud}
\begin{document}

\twocolumn[
\mlsystitle{\srift{}: Swift and Thrifty Distributed Neural Network Training on the Cloud}



\mlsyssetsymbol{equal}{*}

\begin{mlsysauthorlist}
\mlsysauthor{Liang Luo}{uw,fb}
\mlsysauthor{Peter West}{uw}
\mlsysauthor{Pratyush Patel}{uw}
\mlsysauthor{Arvind Krishnamurthy}{uw}
\mlsysauthor{Luis Ceze}{uw,octoml}
\end{mlsysauthorlist}

\mlsysaffiliation{fb}{Currently employed at Meta Platforms Inc.}
\mlsysaffiliation{uw}{University of Washington}
\mlsysaffiliation{octoml}{OctoML Inc.}
\mlsyscorrespondingauthor{Liang Luo}{liangluo@cs.washington.edu}

\mlsyskeywords{Machine Learning, MLSys}
\vskip 0.3in

\begin{abstract}
Finding the best VM configuration is key to achieve lower cost and higher throughput, two primary concerns in cloud-based distributed neural network (NN) training today. Optimal VM selection that meets user constraints requires efficiently navigating a large search space while controlling for the performance variance associated with sharing cloud instances and networks.

In this work, we characterize this variance in the context of distributed NN training and present results of a comprehensive throughput and cost-efficiency study we conducted across a wide array of instances to prune for the optimal VM search space. Using insights from these studies, we built \srift, a system that combines runtime profiling with learned performance models to accurately predict training performance and find the best VM choice that satisfies user constraints, potentially leveraging both heterogeneous setups and spot instances. 
We integrated \srift with PyTorch and evaluated it on Amazon EC2. We conducted a large-scale generalization study of \srift across more than 2K training setups on \ectwo. Our results show that \srift achieves an iteration latency prediction error of 8\%, and its VM instance recommendations offer significant throughput gain and cost reduction while satisfying user constraints compared to existing solutions in complex, real-world scenarios.
\end{abstract}
]



\printAffiliationsAndNotice{}  
\section{Introduction}
To date, most efforts in datacenter and cloud environments focus on improving NN training throughput~\cite{luo2020plink, narayanan2020heterogeneity, thorpe2021dorylus, mudigere2021softwarehardware}. 
However, with the cost of cloud-based NN training soaring to millions of dollars~\cite{OpenAIla69:online}, cost has become another critical concern~\cite{HowdoMLP36:online}.

Finding optimal  VM instances is key to high-throughput, low-cost training. However, given a training job, a time, and a cost constraint, which VM configurations finish the job fastest? Which achieve the lowest cost?


Answering such questions requires accurate estimations of training performance for potentially unseen NN models in a plethora of cloud-provided configurations. Prior work has proposed model-~\cite{8622396,qi2016paleo,peng2018optimus,zheng2019cynthia} and profile-based~\cite{201567, yi2020not,daydream,vanir} techniques for performance prediction, but they fall short of tackling the problems arising from modern cloud:  performance variations introduced by multi-tenancy and the dynamic nature of the network make prediction difficult, especially in the synchronous data parallel training paradigm; the constant billing of VMs and expensive GPU instances limit profiling and exploration; heterogeneous configurations and spot instances might be needed to optimally achieve user objectives; and volatility and interference on cloud resources may require users to continuously revise selected configurations to meet their goals. 



In this work, we present \srift, a system that finds the best VM instances to train an NN model in the cloud given user objectives and constraints. \srift combines model- and profile-based approaches using learned models, lightweight instrumentation, simulation, and hybrid constraint solving to tackle the challenges. It carefully characterizes the temporal and spatial variance induced by the cloud on the compute and communication performance of the distributed training workload; it then uses these empirical measurements to learn performance models that explicitly capture the variance and simulations to accurately predict training iteration latency. \srift leverages  insights from a comprehensive throughput and cost-efficiency study we conducted to trim a large search space that involves heterogeneous and spot VMs  before converting the constraints and goals into a formulation that can be solved. Finally, \srift continually monitors training progress to recommend new VM configurations if large interference or service interrupts violate user constraints.



\noindent
This paper makes the following contributions:

\begin{itemize}[leftmargin=*,noitemsep,topsep=0pt,parsep=0pt,partopsep=0pt,leftmargin=1em]
\item We show why existing solutions fall short of robustly finding the optimal VM configuration given an NN training task by explicitly quantifying the compute and communication performance variance in the public cloud (\S\ref{sec:background}).

\item We present a comprehensive throughput and cost-efficiency study (\S\ref{sec:optimalvmselection}) of training representative NNs on different VM families, sizes, and generations in the cloud to obtain insights needed to prune the search space.

\item We designed and implemented \srift, a system that uses profiling, learned performance models, simulation, and constraint solving to search for the best VM configuration.  Our approach accounts for performance variance, identifies heterogeneous configurations, and takes advantage of spot instances, if necessary, to continually optimize for cost or throughput while meeting user constraints (\S\ref{sec:design}).

\item We integrated \srift with PyTorch and conducted a large-scale generalization study of \srift across more than 2K training setups on \ectwo. In this study, \srift achieves a prediction error of 8\% and finds choices that delivers significantly better throughput and lower cost in real-world training scenarios compared to existing solutions while satisfying user constraints(\S\ref{sec:eval}).

\end{itemize}


\section{Challenges in Cloud-based Distributed Training}
\label{sec:background}


In this work, we focus on synchronous data parallelism using \textit{collective allreduce}~\cite{collectivesOptimization} due to its 
better reproducibility, convergence and performance~\cite{mudigere2021softwarehardware, 10.1145/3458817.3476209, rajbhandari2019zero, lepikhin2020gshard, kumar2021exploring}. We now describe the unique challenges faced in enabling efficient cloud-based distributed NN training.


\subsection{Large VM Selection Search Space}
The cloud creates a large configuration space for distributed training of neural networks. For example, given a global batch size as an input, a user can choose any number of VM instances to distribute the global batch without affecting accuracy; when we factor in user constraints, the decision space further includes heterogeneous and spot VM instances ~\cite{mahgoub2020optimuscloud} in case no single instance type satisfies both time and cost constraints.

\subsection{High Variance in the Cloud Environment}
\label{sec:cloudvariance}
Clouds make it hard to predict iteration latency since shared hardware is not interference-free~\cite{fu2021use}, and we can empirically observe significant spatial (across VMs) and temporal variation in iteration time for the same workload. 

\begin{figure}[t]
	\centering
	\footnotesize
	\includegraphics[width=\linewidth]{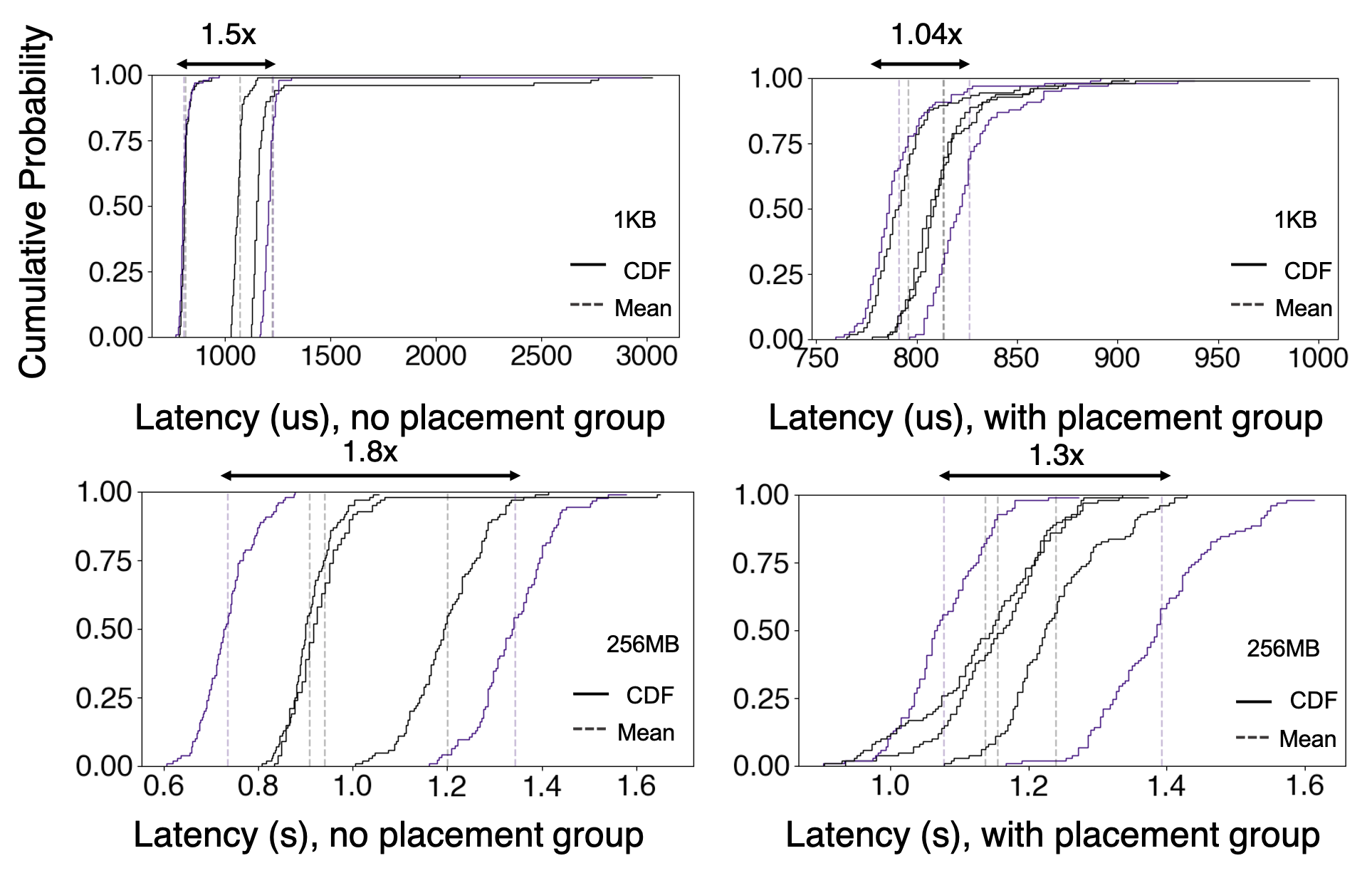}
	\caption{Allreduce latency varies dramatically across both time (up to 2x) and different VM allocations (up to 1.8x).} 
	\label{fig:ec2_allreduce_cdf}
\end{figure}

\noindent\textbf{Communication Variance.} Prior work has observed that the communication performance of VMs varies greatly in the datacenter environment due to oversubscription~\cite{Bilal2012ACS}, multi-tiered topology~\cite{incbricks}, sharing~\cite{luo2020plink}, fairness mechanisms~\cite{Memoryop3:online}, and bursting~\cite{Benchmar28:online}. 
To quantify this variance, we requested 10 allocations of 32 g3.4xlarge instances on \ectwo, 5 without (left) and 5 with (right) placement groups~\cite{aws_placement_groups}, as shown on  Figure~\ref{fig:ec2_allreduce_cdf}; each line represents an allocation. For each allocation, we ran 100 allreduce jobs with NCCL on 1KB (top) and 256MB (bottom) buffers. We find that the mean performance varies up to 1.8x and 1.5x on large and small buffers, respectively. 

\begin{figure}[t]
	\centering
	\footnotesize
	\includegraphics[width=\linewidth]{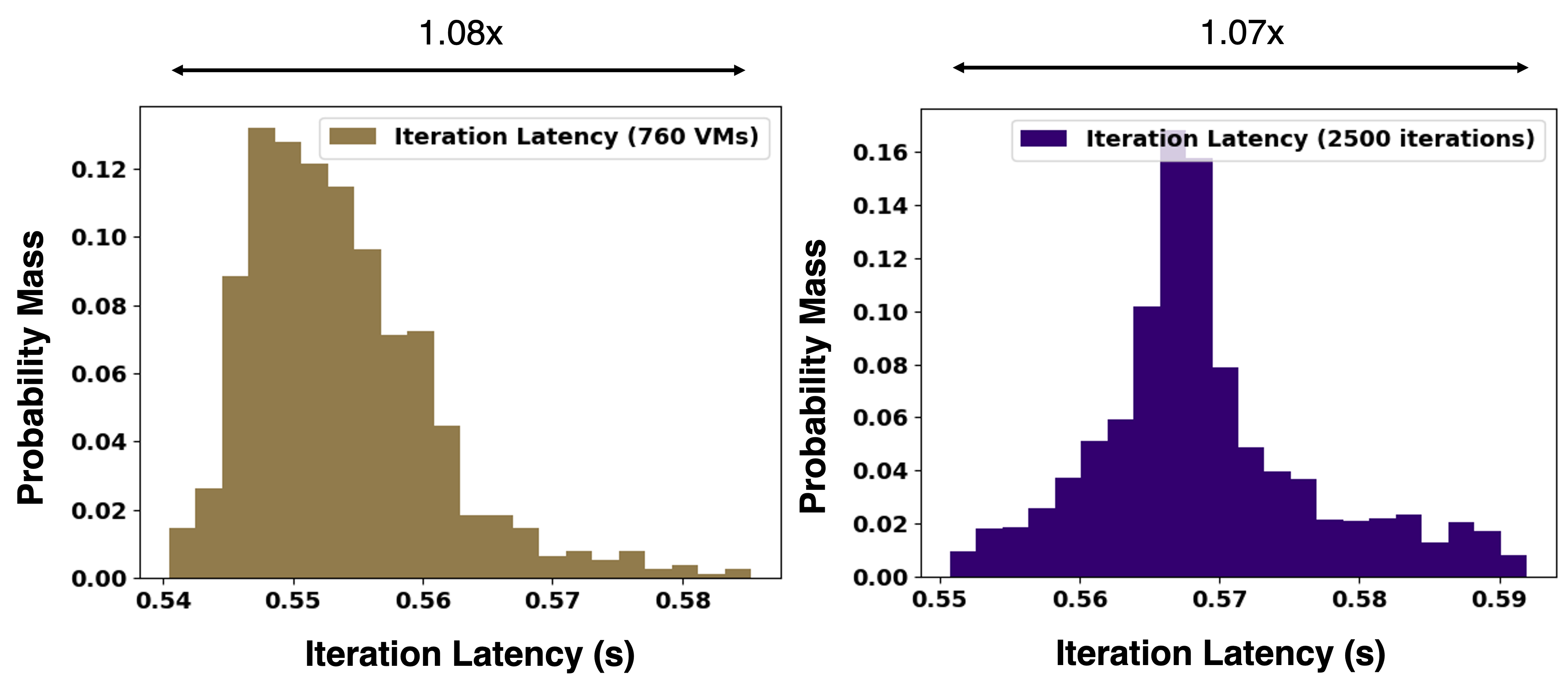}
	\caption{Histogram of compute latency across different VMs (left) and different iterations (right), showing up to a 1.1x variance.
	} 
	\label{fig:compute_variance}
\end{figure}

\noindent\textbf{Compute Variance.} Compute variance has compounding effects: a delay in compute time delays communication and iteration latency, which are determined by the slowest GPU. To characterize spatial variance, we ran independent ResNet18 training tasks on 760 g4dn.2xl instances on \ectwo with a batch size of 64. We plotted the 100-iteration average latency of each VM in Figure~\ref{fig:compute_variance} (left). For temporal variance, we show per-iteration latency across 2500 iterations of a single instance in Figure~\ref{fig:compute_variance} (right). Both histograms resemble normal distributions, with a variance of up to 1.1x.

\subsection{Ineffectiveness of Existing Approaches}
Existing work on selecting appropriate cloud configurations fall into two categories: model~\cite{8622396,qi2016paleo,peng2018optimus,zheng2019cynthia,cai2017neuralpower,8713989,mahgoub2020optimuscloud} or profile based~\cite{vanir, yi2020not,201567,daydream,yadwadkar2017selecting,rubberband}. Most work focuses on \textit{making only homogeneous VM choices while optimizing for a single objective}. 

\noindent\textbf{Model-based solutions} create performance models for various stages of the distributed training process. They often ignore large cloud-induced variance (e.g.,~\cite{qi2016paleo, zheng2019cynthia}) and overlapping between communication and computation~\cite{8622396,10.1145/3341301.3359642,prioritybased,hashemi2018tictac}. 





\noindent\textbf{Profile-based solutions} directly measures specific configurations in the entire search space. To help guide the probes and improve reusability across workloads, D-optimal design, decision forests~\cite{mahgoub2020optimuscloud}, Bayesian Optimization (BO)~\cite{201567}, and workload fingerprinting~\cite{yadwadkar2017selecting} are proposed. Unfortunately, these techniques do not fully address the drawbacks of profile-based solutions because they (1) still incur a high cost due to the need to probe large amounts of expensive VMs, and (2) suffer from unstable measurements due to cloud variance, which mandates repeated probes. 




\label{sec:modelvsprofilebased}
\begin{figure*}
	\centering
	\footnotesize
    	\includegraphics[width=\linewidth]{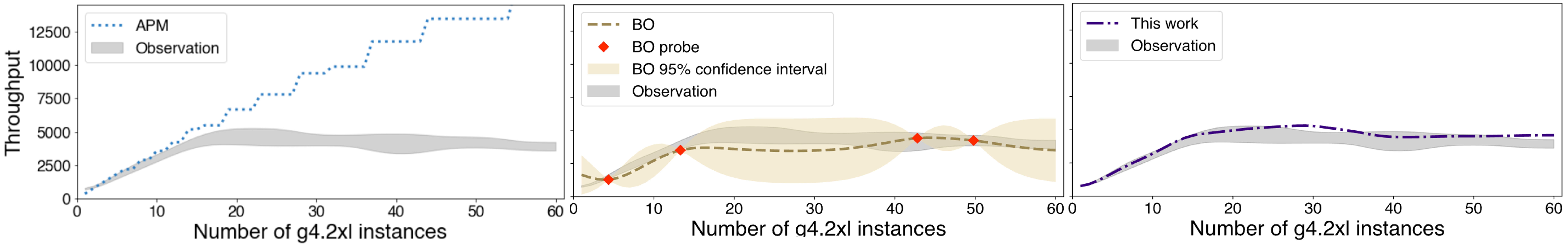}
	\caption{Prediction of APM (left), BO (mid), and \srift (right) on the training performance of ResNet18 on up to 60 nodes on \ectwo instances. Neither APM nor BO finds the best configuration robustly, while \srift achieves high accuracy in the presence of variance.} 
	\label{fig:perf_overview}
\end{figure*}

\textbf{Case Study.}
Even in a homogeneous VM setup, existing approaches are nonoptimal in the cloud environment. To show this, we implemented a model-based prototype, called APM, that combines the compute latency model in Nexus~\cite{shen2019nexus}, the communication latency model in Daydream~\cite{daydream}, and the iteration model in Cynthia~\cite{zheng2019cynthia}. We also built a profile-based prototype based on BO (used by Cherrypick~\cite{201567} and HeterBO). We predicted training throughput of ResNet18 on 60 g4dn.2xl VMs in us-east-1 region of \ectwo with a global batch size of 480, equally distributed to each instance. This training job had 18 possible configurations. We plotted the range of observed and predicted throughput of a given approach versus number of GPUs for each configuration across 7 VM allocations.\footnote{The BO-based model 
is limited to probe 4 times within the first allocation. If BO proposes an invalid VM count, the throughput of the closest observation is used.} 
Figure~\ref{fig:perf_overview} shows the results. Neither APM nor the BO-based solution finds the optimal VM configuration for this workload. APM ignores performance degradation due to the increase in scale and hence exaggerates performance, and its choice is inferior to the optimal by up to 1.1x and 3.2x throughput and cost, respectively. BO's prediction accuracy is negatively affected by both the allocation variance (up to 1.3x) and the number of probes, and its choice is up to 1.2x and 1.7x inferior to the optimal choice throughput- and cost-wise. Further, this non-concave throughput curve causes specific prior-based BOs (e.g., HeterBO) to prematurely stop exploring. 

\begin{table}
	\centering
	\small
	\resizebox{.95\columnwidth}{!}
	{
    	\begin{tabular}{|c|c|c|c|c|c|c|}
    	\hline
    	      & p3.2xl & p3.8xl & g3.4xl & g4dn.4xl & c5.4xl & c5.18xl  \\
    	    \hline
    	    Device & V100 & 4V100 & M60 & T4 & 36 cores & 72 cores \\
    	    \hline
    	    Gbps. & 10* & 10 & 10* & 25* & 10* & 25 \\
    	    \hline
    	\end{tabular}
	}
	\caption{VM specifications used in the study (* indicates up to).} 
	\label{table:nav_config}
\end{table}

\section{Trimming the VM Search Space}
\label{sec:optimalvmselection}
\label{sec:characterization}

\srift aims to find the optimal VM configuration in a \textit{large search space involving heterogeneous VM types, local batch sizes, and billing types}, which necessitates trimming the search space. We do so by conducting a comprehensive performance and cost-efficiency (throughput-per-hour price) characterization on \ectwo using PyTorch.

We benchmarked four models -- ResNet50, Vgg19, SqueezeNet and AlexNet -- for their unique characteristics in terms of compute and communication intensity. We used on-demand price to compute cost-efficiency and report the harmonic mean of cost-efficiency value across them. 
We experimented on 6 representative VM types, shown on Table~\ref{table:nav_config}, each with up to 32 instances. We summarize our findings in the following takeaways, which we then used extensively in the design of \srift (described in the next section).


\begin{figure}[t]
	\centering
	\includegraphics[width=.95\linewidth, trim=3 3 3 0,clip]{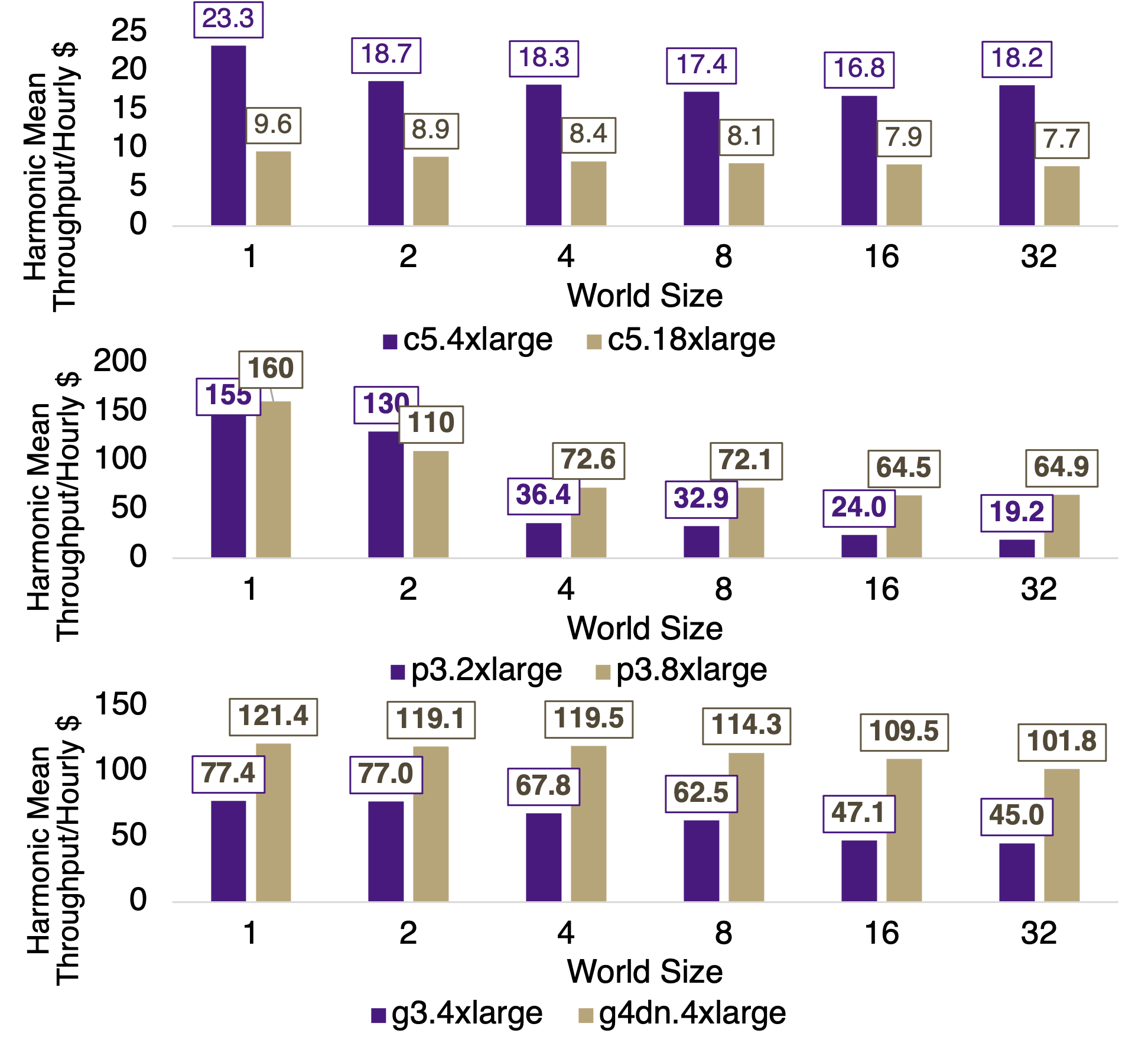}
	\caption{Harmonic mean of the cost-efficiency across 4 different models with varying setups.}
	\label{fig:size_matters}
\end{figure}

\begin{figure}[ht!]
	\centering
	\includegraphics[width=0.95\linewidth]{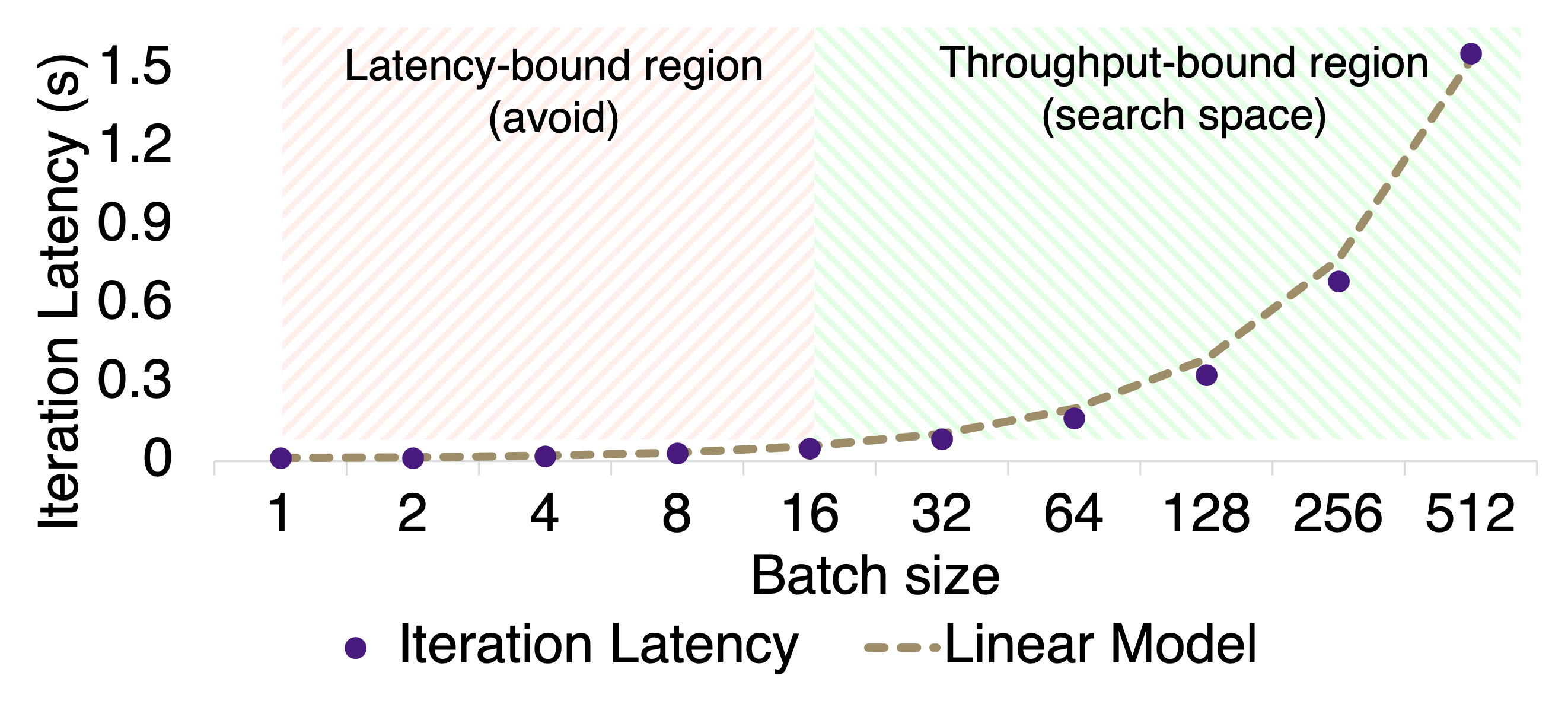}
	\caption{Resnet18 throughput vs batch size (Tesla T4 GPU).} 
	\label{fig:batch_vs_throughput}
\end{figure}

\begin{figure}[ht!]
	\centering
	\includegraphics[width=.95\linewidth]{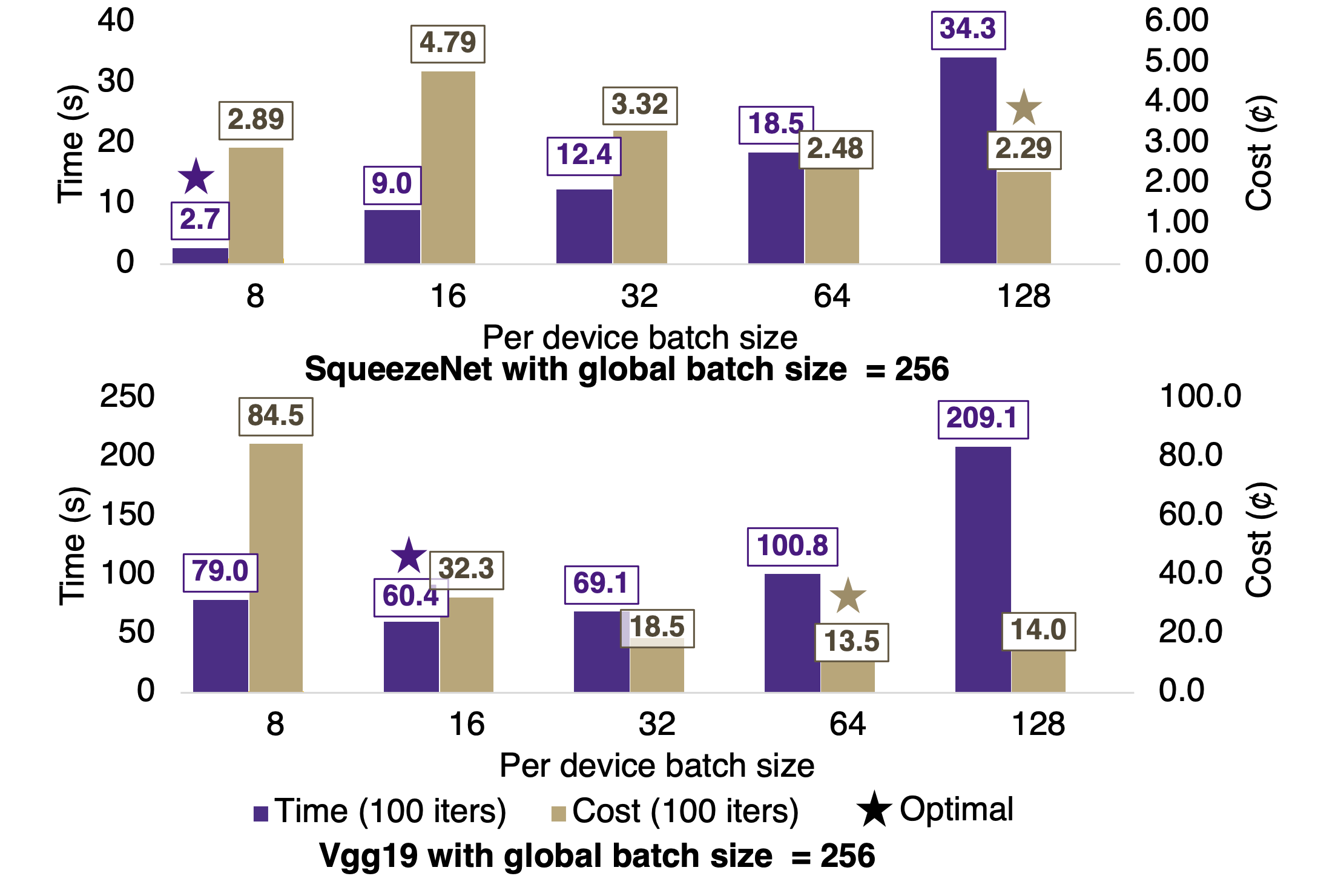}
	\caption{No single 'magic' per-device batch size achieves the best throughput or the lowest cost for all models.}
	\label{fig:cost_throughput_vs_batch_suze}
\end{figure}

\noindent\textbf{Takeaway 1: Prefer GPU over CPU Instances.}
With current pricing, CPUs are still inferior to GPUs in terms of cost-efficiency: Figure~\ref{fig:size_matters} (top and mid) shows that even the least cost-efficient GPU instances outperform the most cost-efficient CPU instances. 

\noindent\textbf{Takeaway 2: Prefer larger GPU instances and smaller CPU instances.}
VMs are priced proportional to their compute capacity. Ideal vertical scaling thus implies that cost-efficiency is constant within the same VM family. In reality, c5 (CPU instance) throughput scales poorly with added CPU cores~\cite{ompoverhead}, and larger p3 (GPU) instances scale near-linearly with added GPUs and additional bandwidth provisioned for larger instances. Thus, we prefer larger GPU, but not CPU, instances.

\noindent\textbf{Takeaway 3: VM generation is not a pruning factor.}
The most recent generations of VMs are not always the optimal choice: g3, p3, and g4dn instances have increasingly more modern GPUs, but none is strictly more powerful and cost-efficient than others (Figure~\ref{fig:size_matters}). 

\noindent\textbf{Takeaway 4: Avoid small local (per-device) batch sizes. }
NN training is latency bound and transitions to a throughput-bound process as batch size increases (Figure~\ref{fig:batch_vs_throughput}). Finding the transition boundary lets us prune the search space to avoid too small of a per-device batch size.

\noindent\textbf{Takeaway 5: World size (number of GPUs) is critical.}
Figure~\ref{fig:cost_throughput_vs_batch_suze} plots the performance for different world sizes given a global batch size of 256. World size has significant implications for both NN training throughput (\textit{12x}) and cost (\textit{6x}). Therefore, the optimal world size must be searched in conjunction with the per-device batch size to arrive at the cost-efficiency sweet spot.

\section{Design and Implementation}
\label{sec:design}

\begin{table*}[t!h!]
	\centering
	\footnotesize
	\resizebox{\linewidth}{!}
	{
    	\begin{tabular}{c c c c c c}
    	\hline
    	      & Variance modeling & Heterogeneous VMs & User objective and constraints & Search cost & Continual optimization \\
    	     \hline
    	     Paleo~\cite{qi2016paleo} & \xmark &  \xmark & Time & Low & \xmark \\
    	     Cherrypick~\cite{201567} &  \textcolor{green}{\cmark} &  \xmark & Time & Higher & \xmark \\
    	     Cynthia~\cite{zheng2019cynthia} & \xmark & CPU instances only & Time and Cost & Low & \xmark \\
    	     OptimusCloud~\cite{mahgoub2020optimuscloud} & \xmark & \textcolor{green}{\cmark} & Cost-efficiency & Budget & \textcolor{green}{\cmark} \\
    	     HeterBO~\cite{yi2020not} & \xmark & \xmark & Time or Cost & High & \xmark \\
    	     \textbf{\srift} & \textcolor{green}{\cmark} & \textcolor{green}{\cmark} & Time and Cost & Low & \textcolor{green}{\cmark} \\ 
    	    \hline
    	\end{tabular}
	}
	\caption{\srift is designed to continually find the best VM configuration that satisfies user constraints at low cost in the presence of cloud variance, leveraging heterogeneous VM cluster and spot instances.} 
	\label{table:qualitative_comparison}
\end{table*}

To effectively determine the optimal VM configuration given a set of user goals and constraints, we require our system to (1) explicitly model cloud variances and provide accurate and robust performance estimations, (2) accurately model modern framework optimizations, such as overlapped communication and computation, (3) support reasoning of heterogeneous VM configurations, (4) handle user objective and constraints efficiently, (5) minimize optimization/exploration overheads, and (6) continually optimize VM configuration during interference and spot instance preemption. These requirements set \srift apart from existing systems (Table~\ref{table:qualitative_comparison}). This section details how \srift combines model- and profile-based approaches, using learned models, lightweight instrumentation, simulation, and constraint solving to meet the requirements. \srift takes as input an NN model, a target global batch size, a user VM quota, and time and cost constraints, and it outputs the chosen VM types, their quantities, and the local batch size assigned to them. 




\subsection{Compute Latency and Gradient Exchange Timestamp Modeling}
\label{sec:compute_latency_models}

Given a model and a batch size, \srift needs to learn a latency vs batch-size model and capture the \textit{gradient exchange timestamps} on the backward pass when allreduce operations are issued to properly model overlapping. 

\textbf{Latency Model.} Traditionally, training latency is obtained with learned performance models ~\cite{qi2016paleo}, with the drawback of requiring NN topology and accurate modeling of peak GPU performance~\cite{zhu2018tbd} or with tracing ~\cite{gujarati2020serving}. 
\srift draws on prior work, which observes that the relationship between runtime latency $t(B)$, given a batch size $B$, follows a linear model: $t(B) = \alpha B + \beta$ (where $\alpha$ and $\beta$ are parameters)~\cite{10.5555/3154630.3154681, shen2019nexus}. Since models exhibit different slopes at different batches sizes ~(Figure~\ref{fig:batch_vs_throughput}), \srift uses a piece-wise linear model. It binary searches the maximum batch size $B_{max}$ that can fit on a device and then tries to capture latency at various batch sizes,  
finding the batch size that transitions from latency to throughput-bound training. 

\textbf{Gradient Exchange Timestamps.}
Modern frameworks overlap parameter exchange and backward pass. Capturing the overlapping behavior requires exact timestamps of when a layer's backward pass completes. \srift  uses lightweight instrumentation through hooks~\cite{nnpackag72:online,tfRegist55:online,sksq96py74:online} to record the timestamp at which each layer finishes back-propagation (i.e., starts allreduce). Timespans are normalized over the total backward pass time to allow extrapolation to different batch sizes. With this instrumentation, \srift can also accommodate model optimizations that change the order of allreduces~\cite{hashemi2020caramel}. This process, done once per model and occurring in parallel on all available GPU types, is practical since only a few GPU models reside in  public clouds. 

\srift explicitly tracks latency deviation during this instrumentation stage for later use in the simulator to construct a probabilistic latency model.

\subsection{A Learned AllReduce Performance Model}
Since users have no visibility or control over the placement of VMs, and local observations suffer from variance and may not represent the whole distribution, it is difficult to derive a mathematical model for communication performance\footnote{Bandwidth rated by cloud providers has a 4000\%+ MAPE (mean absolute percentage error).}. We thus opt to learn an end-to-end allreduce model. 

\textbf{Model Output.} Since we cannot directly predict allreduce latency $t$ (given a gradient size of $s$ and world size $n$) because we need to properly model concurrent communication, \srift needs to predict allreduce bandwidth. 

Instead of predicting algorithm bandwidth ($b_{algo} = \frac{s}{t}$), we opted to predict a different label, the \textit{bus bandwidth} $b_{bus} = \frac{2s(n-1)}{nt}$~\cite{daydream}. Bus bandwidth is the average bandwidth as if physically measured from a network interface during an allreduce operation. Compared to algorithm bandwidth, bus bandwidth incorporates $n$ and $s$, two of the most important features, and reduces aliasing of different setups into the same label. For example, allreduce operations on different $s$ may take the same time to finish, resulting in the same algorithm bandwidth, but they may have different $b_{bus}$. Reducing alias in the dataset helps our model better capture the importance of each feature. 

\textbf{Dataset.} \srift performs a grid probe of allreduce bandwidth by sweeping buffer size from 4B to 512MB and world size from 2 to 64 on g3, g4dn, and p3 instance families of different sizes on \ectwo. This approach also captures link differences (e.g., allreduce via NVLink is captured with a small world size). We included the following handpicked features in our dataset: location (cloud, region, and availability zone), GPU and CPU, rated network throughput by the provider, buffer size, world size, and the number of asynchronous transfers; the dataset contains 40K entries, covering both \ectwo{}'s us-west-2 and us-east-1 regions. We repeated experiments with different VM allocations to capture the variance distribution induced by physical placement and dynamic interference. Our focus on synchronous training prevents the need to measure bandwidth of mixed instance types  because the slowest instance's bandwidth determines the global achieved bandwidth; hence, we need only measure individual instance types.


\textbf{Training. }
We trained regression models for allreduce performance using XGBoost~\cite{chen2015xgboost}. We found that the models predicted negative bandwidth for small buffers when trained on the entire dataset, causing a large test error. We mitigated this by training two models for the dataset, one for $s \le MTU$ and one for $s \ge MTU$, where $MTU$ is the maximum number of bytes a single network packet can carry (9K bytes on \ectwo~\cite{JumboFrame:online}). We performed model selection based on an autotuner and mean absolute percentage error (MAPE), sweeping various hyperparameters, such as objectives including pseudo huber loss~\cite{huber1992robust}, which helps identify outliers. Many frameworks dynamically switch among allreduce implementations based on transfer characteristics, and our grid probe captures this. 

\textbf{Variance Comprehension.} Consider a series of observations on the same configuration ($X$) that have different bus bandwidths ($Y_{i}s$) due to variance. When fit with a loss objective, the learner would explicitly return predictions that robustly minimize overall loss across all observations. 


\textbf{Model Updating.} Given network upgrades that affect performance, the \srift allreduce model must be updated. This involves simply capturing new probes, decaying the weights of stale samples, and retraining.



\subsection{Iteration Simulator}
\label{sec:iteration_simulator}
The simulator predicts the mean iteration latency for a given NN model $M$: $t_{iter}=SIM(M,counts,batches,iters)$ by combining the compute and allreduce performance models. An iteration with batch size $B$ takes $t_{iter} = t_{fw}(B) + max(t_{bw}(B), t_{pe})$ to finish, where $t_{fw}$ and $t_{bw}$ are the latencies for the forward and backward passes, respectively, and $t_{pe}$ is the duration of parameter exchange. 

We first describe how \srift estimates $t_{fw}$ and $t_{bw}$ in the presence of cloud variance and heterogeneous VMs. In synchronous training, $t_{fw}$ is bounded by the slowest VM, which is determined probabilistically: for each chosen VM, \srift samples $iter$ values from a normal distribution (\S\ref{sec:cloudvariance}) fitted with a mean equal to the raw prediction (\S\ref{sec:compute_latency_models}) and a scale set to the standard deviation observed during profiling. The highest sampled latency value for each iteration across all VMs becomes the predicted latency for that iteration. The mean compute latency is then determined by averaging the predicted latency across all iterations.

Next, we used the allreduce bandwidth model to derive $t_{pe}$ with a simulator. The simulator begins at the start of the backward pass (timestamp 0). It tracks an event queue ordered by timestamp: for each NN layer, the simulator enqueues the allreduce transfer start time as an event \code{(start, timestamp)}, where \code{timestamp} is collected through the backward pass instrumentation. The simulator dequeues events from the queue continuously in timestamp order and calculates \code{timespan}, the duration between the current timestamp and that of the previous event. When a \code{start} event is dequeued, the allreduce operation at that layer begins, and a concurrency counter $c$ is incremented. To estimate the transfer bandwidth ($b_{tra}$), the simulator queries the allreduce bandwidth model for the bus bandwidth $b_{bus}$ given the layer size $s$. 
We then add the returned $b_{bus}$ to an aggregate bandwidth counter $b_{agg}$, which represents the total concurrent bandwidth sum for all allreduces. Since each VM instance has a limited total bus bandwidth $b_{cap}$, the simulator allocates total bandwidth to each transfer fairly: $b_{tra} = b_{bus} \text{ if } b_{agg} < b_{cap} \text{, else } b_{tra} = \text{min}(b_{bus}, \frac{b_{cap}}{c})$. 



Using $b_{tra}$, the simulator computes and queues a finish timestamp for each layer. Whenever any event is processed, it updates the estimated finish time using the current \code{timespan}. If the resulting event causes any transfer bandwidth to change, all active operations' estimated finish times are recomputed, and new events are queued. The simulation finishes when no further event is in the queue, and the end timestamp is assigned $t_{pe}$. 



\subsection{\srift Optimizer}
Given a model $M$, global batch size $B_{global}$, number of iterations $N$,  VM instances $0...I$ (spot or on-demand), their user quotas $CAPS[]$ and prices $P[]$, together with probed minimum batch size $thresholds$ (Takeaway 4, \S{\ref{sec:characterization}}), and GPU memory capacity $memcap$, subject to a time constraint $T_{lim}$ and monetary budget $\$_{lim}$, the \srift optimizer searches for  configurations that minimize:
\begin{gather*}
\label{eq:optimizerObj}
   \text{\small A cost or time objective} \\
   \fbox{\resizebox{.6\hsize}{!}{$\quad Nt_{iter}\sum_{i\text{ in } I}{(counts[i]P[i])} \quad \textbf{or} \quad N{t_{iter}}$,}}
\end{gather*}

subject to the following constraints:
\begin{gather*}
    \text{\small 1. Per-VM batch constraints:} \\
    \fbox{\resizebox{.5\hsize}{!}{$\sum_{i\text{ in } I}batches[i]counts[i]=B_{global}$}} \\
    \fbox{\resizebox{.7\hsize}{!}{$\forall_{i \text{ in } I} thresholds[i] \leq batches[i] \leq memcap[i]$}} \\
    \text{\small 2. VM count constraints:} \\
    \fbox{\resizebox{.4\hsize}{!}{$\forall_{i \text{ in } I} counts[i] \leq CAPS[i]$}} \\
    \text{\small 3.  Time or cost constraints:} \\
    \fbox{\resizebox{.9\hsize}{!}{$Nt_{iter} \leq T_{lim}~\textbf{or/and}~Nt_{iter}\sum_{i\text{ in } I}{(counts[i]P[i])} \leq \$_{lim}$}}
\end{gather*}

The output \textit{counts[i]} then stores the number type \textit{i}-th VM in the solution, and $batches[i]$ stores the batch size allocated to each VM of type \textit{i}. $t_{iter}$ in the simulator response. 

Directly encoding $SIM$ into SMT logic would take too long to solve since each exploration results in a simulator invocation. To practically solve this constrained optimization problem, \srift uses a hybrid strategy that prunes the search space before performing an exhaustive search and relies on SMT with approximated constraints if needed.

\textbf{Hybrid Solving Strategy.} \srift begins by pruning the search space using the insights from \S\ref{sec:characterization}: (1) global batch size (and hence local batch size) is usually a power of 2 on GPUs~\cite{machinel38:online,CIFAR10C0:online} to fully utilize GPUs; (2) local batch size should be large enough to saturate the compute capacity; and (3) all instances of the same type should have the same batch size for maximum throughput. \footnote{Otherwise, equally distribute the batch to each instance of that type, and the new throughput is no worse.} 
These let \srift reason about instance types rather than individual instances, reducing the problem complexity to O($(logB_{\textit{global}})^I\prod_{i \text{ in } I} CAPS[i])$. Then, if the reduced problem size is feasible (empirically, $<$ 10k invocations to the $SIM$ routine), \srift performs an exhaustive search. 

Though an exhaustive search is feasible for most practical problems, if the search space is still too large, \srift switches to an approximation scheme to lower the problem into an ILP encoding. Since the iteration latency is bound by the slowest instance, the optimal solution is likely to assign batch sizes to different instances so that compute latencies across all selected instances are approximately the same. Thus, we can sweep the batch size that is the slowest across all GPU types, called a $B_{\text{anchor}}$, and use it as the target iteration latency. With $B_{\text{anchor}}$ set, \srift can compute $batches[]$ for all instances efficiently using a binary search. \srift then queries the solver for an optimal solution to a revised optimization problem, with a proxy goal of minimizing $\sum_{i \text { in } I}counts[i]$ subject to the same constraints. 
For each $B_{\text{anchor}}$, \srift queries the simulator for $Nt_{iter}$. It finally outputs the best throughput or lowest cost configurations across all $B_{\text{anchor}}s$ per user goals.

\subsection{\srift Runtime}
\srift{} supports the use of heterogeneous VMs without affecting model quality, monitors training progress, and reacts to potential service interruptions to enable continual VM configuration optimization. 

\textbf{Model Quality.}
\label{sec:convergence}
Consider the gradient term $g_{i,j}$ produced by the $j$th sample on instance type $i$ in a synchronous data parallelism setting; the sum of loss term $\sum{l_{i,j}}$ is constant regardless of how the global batch size is distributed. However, special care is needed when computing an average gradient with heterogeneous local batch sizes because frameworks such as Pytorch assume that the local batch size on each device is identical and compute the average gradient as simply $\frac{\sum{l_{i,j}}}{\sum counts}$~\cite{pytorchd50:online}. Thus, using heterogeneous batch sizes causes samples from instances with smaller batch sizes to receive a disproportionally larger weight and hence may have implications on convergence. \srift uses a technique similar to~\cite{ding2020hetseq,chen2020semi,8367216} by reweighting sample gradients produced by each instance with type $i$ with the coefficient $\frac{batches[i]}{B_{global}}$, so that each sample contributes to the averaged gradient term equally. Thus, from the optimizer’s perspective, all GPUs receive a local batch size that equals the average batch size; therefore, \srift has no impact on quality. 

\textbf{Continual Optimization.}
When unexpected variance or VM preemption occurs, the initial VM configuration may not be optimal~\cite{mahgoub2020optimuscloud}. Thus, \srift must continually optimize VM configurations by taking into account current progress against the original constraints. The \srift runtime tracks the current elapsed time $t$, cost $c$, and iterations $n$ finished. If current progress falls behind its original schedule, \srift reruns the optimizer with updated constraints (\code{N-=n}, \code{\$$_{lim}$-=c, T$_{lim}$-=t}) and the original objectives. However, blindly switching to a new configuration may not be efficient if the variance is transient since overheads result from stopping and resuming the current task. \srift thus maintains a windowed throughput and its standard deviation for the 
previous K (a tunable parameter empirically set to 5) minutes, computes the 95\% confidence bound of throughput, and uses this optimistic throughput to evaluate constraint satisfiability. \srift recommends switching to new VM configurations only when it is highly certain that the current configuration will lead to a violation.

The same procedure occurs during VM preemptions. \srift uses preemption as a signal of depletion of that instance type and does not choose that instance again. When a new VM configuration is proposed, \srift relies on cloud-specific mechanisms (e.g., persistent disk~\cite{AmazonEC44:online}) an9d framework-level elasticity functions~\cite{TorchEla89:online,or2020resource} to checkpoint training progress. The overheads of switching to a new configuration (e.g., launching new VMs) are set as parameters to the optimization process. 


\section{Evaluation}
\label{sec:eval}
Our evaluation goals are to: (1) quantify the benefits of \srift{}'s VM proposal and \srift overhead, (2) establish \srift{}'s generalizability, and (3) demonstrate the effectiveness of \srift{}'s continual optimization.

\subsection{Evaluation Setup and Baselines}
We evaluated \srift with PyTorch 1.5 and NCCL 2.4.8 using CUDA 10.1 and CuDNN 7 on Linux kernel 5.3.

We ran experiments on \ectwo{}. Our study can prune most of the thousands of potential instances in \S\ref{sec:characterization}  for our workloads. We selected g3, g4dn, and p3 families of VMs. Note that \srift must still deal with a large search space even after pruning due to heterogeneity and variable local batch sizes.

\begin{table}[t]
	\centering
	\footnotesize
	\resizebox{\columnwidth}{!}
    {
    	\begin{tabular}{|c|c|c|c|}
    	\hline
    	    Network (Abbr) & p3.8xl & g3.8xl & g4dn.2xl/8xl \\
    	    \hline
    	    AlexNet~\cite{alexnet} (ALN) & 8192 & 2048 & 2048 \\
    	    \hline
    	    ResNet18\&50~\cite{resnet} (RN18) & 2048\&512 & 512\&64 & 512\&64/128 \\
    	    \hline
            Vgg16\&19~\cite{vgg19} & 512 & 64 & 128 \\
    	    \hline    	    
    	    ResNext50\_32x4d~\cite{resnext} (RNX) & 512 & 64 & 64 \\
    	    \hline
    	    SqueezeNet\_1\_1~\cite{squeezenet} (SQN) & 2048 & 512 & 512 \\
    	    \hline
    	    ShuffleNet\_v2\_x2\_0~\cite{ShuffleNet} (SFN) & 1024 & 256 & 256 \\
    	    \hline
    	    Inception\_v3~\cite{szegedy2015rethinking} (INC) & 512 & 128 & 128 \\
    	    \hline
    	    BERT-Base-Cased~\cite{BERT} & 256 & 64 & 64 \\
    	    \hline
    	    Xlm-Clm-Ende-1024~\cite{XLM} & 512 & 64 & 128 \\ 
    	    \hline
    	    DLRM~\cite{DLRM19} & 512K & - & 128K \\
    	    \hline
    	    \hline
    	    \textbf{Hourly rate} & \textbf{\$3.68} & \textbf{\$0.69} & \textbf{\$0.72} \\
    	    \hline
    	\end{tabular}
    }
	\vspace{-2ex}	
	\caption{Supported instance prices and max batch sizes.}
	\label{table:eval_config}
\end{table}

We report the average latency of at least 20 iterations and ignore the once-per-DNN profiling time. We use spot instance prices at the time of writing. We include DNNs from vision (synthetic ImageNet dataset), NLP (Huggingface transformers with the built-in dataset), and recommendation models (Facebook DLRM modified to use data parallelism with Criteo TB Click Logs) to evaluate \srift (Table~\ref{table:eval_config}). We use mean absolute percentage error (MAPE) to evaluate \srift{}'s one-shot prediction accuracy. Since \srift has no impact on model quality~(\S\ref{sec:convergence}), we use throughput as the speedup metric. We compare \srift with various baselines by replacing its learned models with the baselines', keeping the optimizer intact for fairness.

\noindent\textbf{Paleo}~\cite{qi2016paleo}, an analytical model-based predictor. We use default settings and added device specifications for relevant GPUs from~\cite{teslam6076:online}. Paleo does not model NCCL performance but instead  estimates for individual allreduce implementations; we thus report the average. 

\noindent{\textbf{APM}}, a model-based solution defined in~\S\ref{sec:optimalvmselection} that combines the linear compute latency model used in Nexus, the allreduce latency model used in Daydream and by Nvidia~\cite{nccltest39:online}, and the iteration model used in Cynthia. We introduce APM as a strong baseline because it shares the \srift optimizer and hence can reason about heterogeneity efficiently.

\noindent{\textbf{Oracle BO}}, an oracle Bayesian Optimizer baseline (used by Cherrypick and HeterBO). We allow exploring of one-third of the possible configurations. This baseline represents the best that any BO-based approach can do since it explores directly on the ground truth; therefore, it is not affected by cloud variance.

\noindent{\textbf{Greedy}}, two widely used greedy heuristics: (1) the \textit{full-batch-size, cheapest GPU first (CGF)} policy, which uses the largest batch size on all chosen GPUs~(Table~\ref{table:eval_config}) and prefers cheaper GPUs, and (2) the \textit{magic-batch-size}~\cite{neuralne31:online, Examples63:online,Whatisth65:online}, \textit{fastest GPU first (FGF)} policy, which  uses a fixed device batch size of 64 and favors faster GPUs. In case of insufficient GPUs, FGF fully packs all GPUs with the largest feasible batch size. 

\subsection{End-to-end Benefits of \srift: Case Studies}
\label{sec:eval_case_study}
We highlight the benefits of \srift{} through case studies, where we evaluate (1) the actual throughput/cost of configurations proposed by \srift and baselines on \ectwo, 
and (2) \srift's overhead. To assess \srift{}'s scalability, we use typical global batch sizes from a few hundreds to thousands as well as a user quota of tens of instances per VM type. We use the format \code{<num><instance>@batch} to represent the use of \code{<num>} quantities of instance type \code{instance}, each with a batch size of \code{batch}.



\noindent\textbf{Goal 1: Maximizing throughput, homogeneous VMs}

\newcounter{mycounter}
\newcommand\showmycounter{\stepcounter{mycounter}\themycounter}

\noindent\textit{User quota:} 64 g4dn.8xl instances. 

\noindent\textit{Case \showmycounter}. Minimize ResNet50 training time. Batch size: 128.
\begin{table}[h!]
	\centering
	\footnotesize
	\resizebox{\columnwidth}{!}
	{
    	\begin{tabular}{|c|c|c|c|c|c|c|}
    	    \hline
    	    ResNet50    & \srift    & O-BO/Paleo  & APM      & FGF      & CGF  \\
    	    \hline  
    	    Config      &  16g4dn@8 & 8g4dn@16   & 32g4dn@4 & 2g4dn@64 & 1g4dn@128\\
    	    \hline
    	    Actual lat. &\textbf{0.13s/iter}  & 0.17s/iter & 0.17s/iter  & 0.58s/iter & 1.1s/iter\\ 
    	    \hline
    	\end{tabular}
	}
\end{table}
\noindent\textit{Explanation:} \srift{} returns in 1.1s. \srift{} assessed all possible configurations and learns that ResNet50 is compute intensive. \srift{} decides to parallelize training on 16 VM instances for optimal throughput.

\noindent\textit{User quota:} 64 g3.8xl instances. 

\noindent\textit{Case \showmycounter}. Minimize training time for Vgg16. Batch size: 512.

\begin{table}[h!]
	\centering
	\footnotesize
	\resizebox{.75\columnwidth}{!}
	{
    	\begin{tabular}{|c|c|c|}
    	    \hline
    	    Vgg16 & \srift & Paleo/APM/FGF/CGF/O-BO\\
    	    \hline
    	    Config & 32g3@16 & 8g3@64  \\
    	    \hline
    	    Actual lat. &\textbf{1.03s/iter} & 1.11s/iter \\ 
    	    \hline
    	\end{tabular}
	}
\end{table}

\noindent\textit{Explanation:} \srift{} returns in 0.8s. Although \srift{} learns that Vgg16 is both compute and communication intensive, it identifies the sweet spot of 32 VMs for a 7\% additional throughput gain, disagreeing with all baselines. 

\noindent\textit{Case \showmycounter}. Minimize AlexNet training time. Batch size: 1K.


\begin{table}[h!]
	\centering
	\footnotesize
	\resizebox{.8\columnwidth}{!}
	{
    	\begin{tabular}{|c|c|c|c|}
    	    \hline
    	    AlexNet & \srift/Paleo/APM & O-BO/FGF & CGF  \\
    	    \hline  
    	    Config &  2g3@512   & 16g3@64 &  1g3@1024 \\
    	    \hline
    	    Actual lat.           &\textbf{0.415s/iter}  & 0.425s/iter & 0.535s/iter \\ 
    	    \hline
    	\end{tabular}
	}
\end{table}

\noindent\textit{Explanation:} \srift{} returns in 0.9s. It learns that the throughput vs world size curve is concave. \srift{} avoids evaluating configurations with per-device batch sizes smaller than 64  since they do not fully utilize GPU capacity.


The following sections address heterogeneous choice of VMs, which is not supported by Paleo and O-BO. We dropped Paleo and limited O-BO's search within homogeneous setups by repeating the search process for each VM type and equally splitting the exploration quota.

\noindent\textbf{Goal 2: Minimizing cost, heterogeneous choices}

\noindent\textit{User quota:} 32 instances each of g4dn.8xl and g3.8xl. 

\noindent\textit{Case \showmycounter}. Minimize DLRM cost. Batch size: 1K.


\begin{table}[h!]
	\centering
	\footnotesize
	\resizebox{.85\columnwidth}{!}
	{
    	\begin{tabular}{|c|c|c|c|>{\columncolor{LightGray}}c|}
    	    \hline
    	    DLRM & \srift/APM/CGF & FGF & O-BO & Paleo\\
    	    \hline
    	    Config & 1g4dn@1024 & 16g4dn@64 & 4gdn@64 &  - \\
    	    \hline
    	    Actual cost.           & \textbf{0.0024¢/iter} &  0.601¢/iter & 0.130¢/iter  & -\\ 
    	    \hline
    	\end{tabular}
	}
\end{table}

\noindent\textit{Explanation:} \srift{} returns in 4.3s. It learns that DLRM running under data parallelism is communication heavy because the embedding tables must be synchronized. \srift agrees with APM and CGF that the best strategy is to pack all batches on the fewest GPUs possible.

\noindent\textit{Case \showmycounter}. Minimize XLM and BERT training cost. Batch size: 512 and 1K.

\begin{table}[h!]
	\centering
	\footnotesize
	\resizebox{.82\columnwidth}{!}
    {
    	\begin{tabular}{|c|c|c|c|>{\columncolor{LightGray}}c|}
    	    \hline
    	    XLM & \srift & APM/CGF & FGF/O-BO & Paleo  \\
    	    \hline
    	    Config & 4g4dn@128 & 8g3@64 & 8g4dn@64 & -  \\
    	    \hline
    	    Actual cost.           &\textbf{0.174¢/iter} & 0.223¢/iter & 0.197¢/iter & - \\ 
    	    \hline
    	\end{tabular} 
    }\\
    \smallskip
	\resizebox{.82\columnwidth}{!}
    {    
    	\begin{tabular}{|c|c|c|c|>{\columncolor{LightGray}}c|}
    	    \hline
    	    BERT & \srift/O-BO/CGF/APM & FGF & Paleo  \\
    	    \hline
    	    Config & 16g3dn@64 & 16g4@64 &  -  \\
    	    \hline
    	    Actual cost.  &\textbf{0.343¢/iter} & 0.416¢/iter &  - \\ 
    	    \hline    	
    	\end{tabular}
    }
\end{table}
\noindent\textit{Explanation:} \srift{} returns in 2.4s (XLM) and 6.5s (BERT). On XLM, \srift learns that using more GPUs with a smaller batch size for a higher degree of parallelism cannot outweigh the overhead of communication, and its choice results in up to a 1.26x better cost. On BERT, all solutions except FGF converge on fully packing 16 g3 instances to save cost.

\noindent\textbf{Goal 3: Minimizing time, heterogeneous choices with constraints}

\noindent\textit{User quota:} 4 p3.8xl, 8 g3.8xl and g4dn.8xl instances.

\noindent\textit{Case \showmycounter:} Train Inception for 500 iterations with a global batch size of 2.1K in 5 minutes and $\$1.4$. Minimize time.

\begin{table}[h!]
	\centering 
	\footnotesize
	\resizebox{\columnwidth}{!}
    {
    	\begin{tabular}{|c|c|c|c|c|>{\columncolor{LightGray}}c|}
    	    \hline
    	    Inception V3 & \srift & APM & FGF & CGF & Paleo/O-BO \\
    	    \hline
    	           & 4p3@512+ & 4p3@512+ & 4p3@512+   & 1p3@512+ & \\
    	    Config & 4g4dn@32 & 8g4dn@16  & 1g4dn@128  & 8g3@128+ & -\\
    	           &          &          &            & 8g4dn@128 & \\
    	    \hline
    	    Actual time & \textbf{259s} & \textbf{260s} & 787s & 787s & -\\ 
    	    \hline
    	    Actual cost & \textbf{\$1.26} & \$1.47 & \$6.74 & \$3.26 & -\\ 
    	    \hline
    	\end{tabular}
    }
\end{table}


\noindent\textit{Case \showmycounter:} Finetune ShuffleNet for 1k iterations with a global batch size of 6K in 6 minutes and $\$2.5$. Minimize time.

\begin{table}[h!]
	\centering
	\footnotesize
	\resizebox{\columnwidth}{!}
    {
    	\begin{tabular}{|c|c|c|c|c|c|>{\columncolor{LightGray}}c|}
    	    \hline
    	    ShuffleNet & \srift & APM & FGF & CGF & Paleo/O-BO \\
    	    \hline
    	           &             & 4p3@1024+ & 4p3@1024+ &2p3@1024+ & \\
    	    Config & UNSAT       & 8g3@128 + & 8g4dn@256 &8g3@256+  & -\\
    	           &             & 8g4dn@128 &           &8g4dn@256 & \\
    	    \hline
    	    Actual time & \textbf{N/A} & 381s & 759s & 761s & -\\ 
    	    \hline
    	    Actual cost & \textbf{N/A} & \$2.74 & \$4.31 & \$3.93 & -\\ 
    	    \hline
    	\end{tabular}
	}
\end{table}

\noindent\textit{Explanation:} \srift returns in 2.5s and 0.9s, respectively. It is forced to make a heterogeneous choice for Inception training because no homogeneous choice can fit the global batch size. \srift makes per-device assignments that roughly balance the computation latency across different instances with local batch size, resulting in lower cost. In the case of ShuffleNet, \srift believes the given constraints are too tight and hence it did not provide any solution. In both cases, other solutions produced configurations that violated user constraints.

In summary, we showed that \srift{}'s VM configurations in complex scenarios with a wide range of models outperform baselines in terms of throughput and/or cost. 

\subsection{\srift Generalizability: Accuracy of Prediction}
\label{sec:ablation_study}
We performed an ablation study of prediction accuracy for the learned compute and communication models and the simulator. We compared \srift{} to the strong APM baseline. 

\textbf{Compute-latency Prediction Accuracy.} We trained on different GPUs and swept batch sizes from 1 to maximum in a geometric sequence with powers of 2; we then measured the iteration latency as ground truth. We limited \srift to probe at 4 different batch sizes, regardless of model, and then we compared the predicted latency versus the ground truth. The results are summarized in Table~\ref{table:compute_acc}. Overall, \srift{}'s compute-latency model achieves a MAPE of 6.4\%, 5.9\%, 4.5\% compared to APM's 12.5\%, 9.4\% and 8.5\% when predicting forward, backward, and the entire iteration latency, respectively. 

\begin{table}[t]
    \centering
    \resizebox{\columnwidth}{!}
    {
        \begin{tabular}{|c|c|c|c|c|c|c|c|c|c|c|c|}
        \hline
            & RNX & SQN & SFN & RN18 & vgg19 & RN50 & INC & ALN & BERT & XLM & DLRM \\
        \hline
        APM & 3.1\% & 19\%& 19\%& 11\%& 5.2\% & 7.3\% & 9.4\% & 4.1\% & 3.9\% & 6.1\% & 1.1\% \\
        \hline
        \srift & 2.3\% & 5.2\% & 8.0\% & 3.1\% & 3.6\% & 3.0\% & 5.4\% & 4.0\% & 12\% & 4.8\% & 1.2\% \\  
        \hline
        \end{tabular}
    }
	\\
	\smallskip    
    \resizebox{.55\columnwidth}{!}
    {
        \begin{tabular}{|c|c|c|c|}
        \hline
            & Tesla M60 & Tesla T4 & Tesla V100  \\
        \hline
        APM & 6.5\% & 7.8 \% & 12\% \\
        \hline
        \srift{} & 3.2\% & 4.4\%  & 6.2\% \\ 
        \hline
        \end{tabular}
    }     
	\caption{Comparison of compute latency models' MAPE aggregated by NN model and by GPU.}
	\label{table:compute_acc}    
\end{table}


\textbf{Allreduce Bandwidth Model Accuracy.} Our model achieves a MAPE of 11.7\% on large transfers (buffer size larger than an MTU) and 23.9\% on small transfers (buffer size no larger than an MTU) in test. The error originates because each configuration (feature) is probed multiple times by reallocating VMs in our dataset, giving different observations (labels) each time. Thus, no model achieves a perfect error rate. Our analysis shows a lower bound on error rate of 9.6\% and 8.2\% for small and large transfers, respectively. Our model's accuracy is close to the best achievable for large transfers; we are less concerned about the higher error rate on small transfers because they translate to only tens of milliseconds of transfer time. 

\textbf{End-to-end Accuracy.} We predicted end-to-end training iteration latency for a large number of real job configurations on EC2; each configuration had different models, batch sizes, or world sizes and was launched on different instance types, regions, availability zones and placement groups, with a total of 2K experiments.\footnote{Due to resource constraints, not all configurations were run on all instance types, regions, availability zones and placement groups since we aimed to cover more configurations. In particular, we evaluated vision models across all selected instances, NLP models on the g3 and g4dn instances, and DLRM on the g4dn instances.} We then let \srift predict the latency of each experiment. To report \srift's MAPE comprehensively and succinctly, we summarize in Table~\ref{table:end_to_end_acc} \srift{}'s high accuracy and ability to generalize across 3 dimensions: model, world size, and instance types. 
Overall, \srift{} achieves a MAPE of 8.3\% versus the  24\% of APM. This confirms the crucial role that gradient time-stamping, the learned performance model, and the simulator play in delivering an accurate prediction. As result, when applying \srift to the same training task in~\S\ref{sec:modelvsprofilebased}, it achieves a much lower prediction error, as shown in Figure~\ref{fig:perf_overview} (right).

\begin{table}
    \centering
    \resizebox{\columnwidth}{!}
    {
        \begin{tabular}{|c|c|c|c|c|c|c|c|c|c|c|c|}
        \hline
         DNN     & RNX    & SQN    & SFN    & RN18  & vgg19  & RN50    & INC    & ALN    & BERT  & XLM   & DLRM \\
        \hline
         APM     & 18\%   & 16\%   & 18\%   & 23\%  & 25\%   & 19\%    & 15\%   & 27\%   & 24\%  & 33\%  & 59\% \\
        \hline
        \srift   & 8.4\%  & 9.9\%  & 6.5\%  & 6.5\% & 9.8\%  & 7.5\%   & 8.1\%  & 12\%   & 6.8\% & 8.2\% & 3.9\% \\ 
        \hline
        \end{tabular}
    }
	\\
	\smallskip
	\resizebox{.7\columnwidth}{!}
    {
        \begin{tabular}{|c|c|c|c|c|c|}
        \hline
        VM type & g3.8xl & p3.8xl & g4dn.8xl & g4dn.4xl* & g4dn.2xl*  \\
        \hline
        APM    &  20\%  & 21\%    & 32\%     &    21\%   & 23\% \\
        \hline
        \srift & 7.3\% & 8.9\%    & 8.1\%    &    8.9\%    & 9.9\% \\ 
        \hline
        \end{tabular}
    }     
	\\
	\smallskip
	\resizebox{.7\columnwidth}{!}
    {
        \begin{tabular}{|c|c|c|c|c|c|c|c|}
        \hline
        VM count & 2 & 4 & 8 & 16 & 32 & 64 \\
        \hline
        APM      & 18\% & 20\% & 21\% & 24\% & 29\% & 33\% \\
        \hline
        \srift   & 8.2\% & 7.0\%  & 7.1\% & 8.0\%  & 9.5\% & 12\% \\ 
        \hline
        \end{tabular}
    }     
	\caption{MAPE of end-to-end predictions aggregated by NN, instance type and VM count. *: This instance has a variable bandwidth.}	\label{table:end_to_end_acc}    
\end{table}

\begin{figure}[t!]
	\centering
	\includegraphics[width=.7\linewidth]{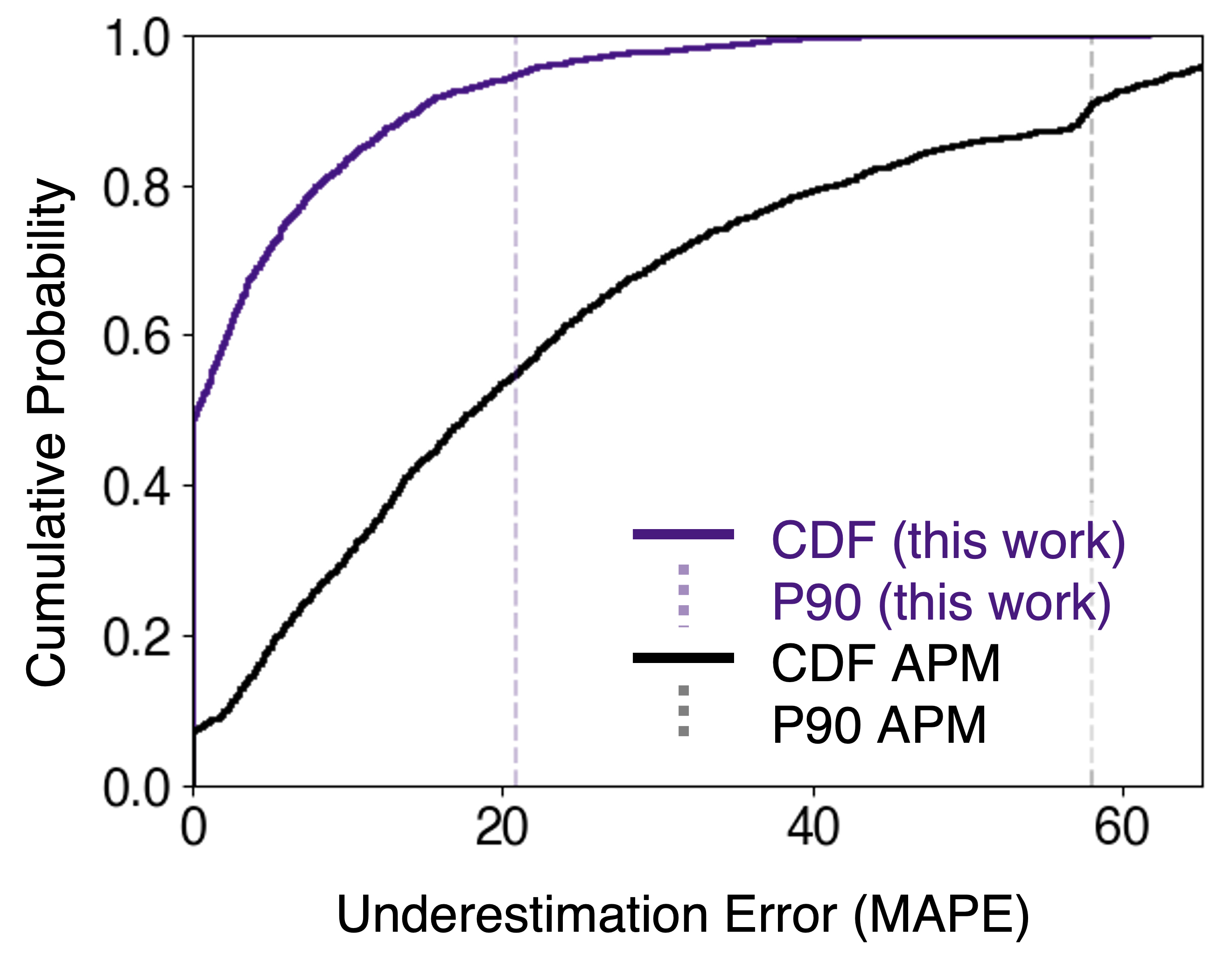}
	\caption{CDF of underestimation error for \srift and APM.}
	\vspace{-1ex}	
	\label{fig:histogram}
\end{figure}

Though \srift performs well overall, underestimation error is more serious than overestimation error since the former can violate user constraints. We now quantify underestimation error in Figure ~\ref{fig:histogram}. In 48\% of the cases \srift does not underestimate; in 90\% of the cases when it does, its MAPE is no more than 21\%. APM, on the other hand, underestimates 93\% of the time, with a 58\% underestimation MAPE. Further analysis shows that most error comes from: (1) allocation variance: we observed up to a 1.15x variance across different allocations; thus, \srift{} cannot achieve a good MAPE on these setups because it is making a one-shot prediction, highlighting the necessity of variance modeling; (2) using a small batch size  since 
it is latency or bandwidth sensitive and more subject to intra-VM and network noise; and (3) using a large world size, which is prone to inter-VM variance and desynchronization.

\subsection{Continual Optimization}
We now evaluate how \srift{}'s runtime continually optimize VM configuration to satisfy original constraints in the event of Spot instance preemption. We set some of the hyperparameters empirically: we expect \ectwo{}'s instance launch time to be 150s and EBS's detach time to be 5s. We give \srift a 1s solving time and set its instance preference list to p3.8xl, g3.8xl then g4dn.8xl, with a user quota of 2, 8, and 8 respectively.

\begin{figure}[t]
	\centering
	\includegraphics[width=\linewidth]{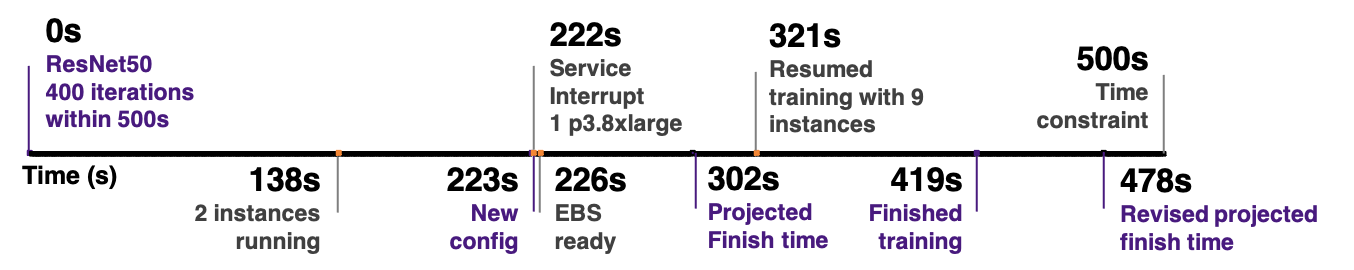}
	\vspace{-5ex}	
	\caption{Timeline of \srift reacting to a service interrupt.}
	\label{fig:reactive_timeline}
\end{figure}

We trained ResNet50 with a batch size of 1K for 400 iterations and a time limit for 500s, with a goal of minimizing cost. \srift started training with 2 p3.8xlarge spot instances, each with a 512 batch size. With this setup, the job was projected to finish in 152s (at time=302s, with 150s for launching instances). In fact, the instances finished launching at 138s.  When 200 iterations completed, at time 222s, we canceled one spot instance to simulate a service interrupt. \srift detected the interrupt immediately at 223s and started working on an alternative configuration while \ectwo made the terminated instance's volume ready, which took 4s. \srift further needed to give \ectwo 150s to boot up an instance. Thus, \srift had to derive a cost-efficient plan that finished in 123s. It proposed the use of an additional 8 g3.8xl (p3.8xl was not available) instances, each with a batch size of 64. The remaining 200 iterations were projected to finish in 101s, at 479s. In fact, the new instances booted in 95s. At time 419s, training completed. We summarize the key events in Figure~\ref{fig:reactive_timeline}: \srift used current progress to update constraints to propose a new configuration of instances, satisfying the original job's constraints.
\section{Related Work and Discussion}
\label{sec:discussion}
\noindent\textbf{Performance Modeling.} Paleo and ~\cite{8713989} use detailed knowledge of the DNN and peak GPU flops to estimate compute latency. Neuralpower~\cite{cai2017neuralpower} draws a correlation between parameter count and runtime. ~\cite{8622396} trains a neural network to infer runtime. They all work only on known NNs and ignore cloud variance (\S\ref{sec:optimalvmselection}). \srift improves on the model used in~\cite{10.5555/3154630.3154681, shen2019nexus,qiao2020pollux} to predict compute latency. For communication latency modelling, Cynthia and Optimus~\cite{peng2018optimus} consider only the parameter server (PS) architecture. Paleo, Optimus and Cynthia rely on accurate bandwidth estimates. \srift uses a learned model that significantly lowers error compared to approaches that rely on an accurate, static bandwidth reading (e.g., Daydream has 34\% error rate predicting allreduce performance). For overlap modelling, Cynthia assumes full overlapping, underestimating iteration time; Paleo and Optimus ignore overlap. Pollux~\cite{qiao2020pollux} learns an overlapping factor during training. \srift collects detailed traces of when layer gradients become available to accurately model overlapping. FlexFlow~\cite{jia2018beyond},~\cite{mirhoseini2017device} and~\cite{rubberband} learn performance models on a predefined cluster and do not suggest VM configurations. In terms of heterogeneity, Pollux does not consider it, which  can fail to find any valid solution. Gavel~\cite{narayanan2020heterogeneity} solves an orthogonal scheduling problem given a known cluster; it supports heterogeneity temporally: jobs run on homogeneous hardware at any given time and can be migrated to different hardware later. Dorylus~\cite{thorpe2021dorylus} focuses on CPU-based, asynchronous GNN training on lambda  and does not consider heterogeneity.

\noindent\textbf{Cost Awareness in Cloud-based DNN Training.} Cynthia predicts the optimal number of worker and PS nodes to minimize cost, with a time constraint for CPU instances only. ~\cite{deepak2020analysis} conducted an analytical study on how to leverage multiple clouds and spot pricing for cost-reduction. Elastic frameworks~\cite{Or2020ResourceEI} can improve cost-efficiency by adjusting training nodes with trial and error but do not assume optimality or deal with constraints directly. Proteus~\cite{proteus} exploits spot instances for PS-based elastic training with a bidding algorithm to cheaply procure transient instances to lower cost, but it does not accept user constraints. FC$^2$~\cite{fc2} shares a goal similar goal to \srift but uses simple optimization heuristics, compromising on the selection objective. Cherrypick and Vanir~\cite{vanir} combine a series of heuristics and ML techniques to optimize cloud-based distributed workloads; HeterBO use a Bayesian Optimization approach with search space pruning to efficiently explore instance selection. They all require more extensive exploration and benchmarking than \srift, leading to potentially higher exploration costs.  

\section{Conclusion}
Finding the best instances to meet user constraints in cloud-based distributed NN training is difficult due to the large search space and challenges from the highly variable cloud environment. We designed and implemented \srift{}, a system that draws insight from a comprehensive throughput and cost-efficiency study we conducted to accurately predict training iteration time; the study also pinpoints the best instance configurations to reduce runtime and cost given constraints, with a low profiling cost. We showed on unmodified EC2, with Pytorch, that \srift{} achieves a low end-to-end prediction error and significantly improves throughput and reduces cost. 

\section*{Acknowledgements}

This work was supported in part by the National Science Foundation under Grants \#1518703 and \#1723352, DARPA FastNICs, and gifts from Intel, VMWare, and Google.

\bibliography{mlsys}
\bibliographystyle{mlsys2022}

\end{document}